\documentclass[prl,twocolumn,showpacs,floatfix,aps]{revtex4}

\usepackage{graphicx}
\begin{document}

\title{MgB$_{2}$: synthesis, sound velocity, and dynamics of the vortex
phase}

\author{T.\ V.\ Ignatova, G.\ A.\ Zvyagina, I.\ G.\ Kolobov, E.\ A.\
Masalitin, and V.\ D.\ Fil'\thanks{E-mail: fil@ilt.kharkov.ua}}

\address{B. Verkin Institute for Low Temperature Physics and
Engineering, National Academy of Sciences of Ukraine,  Lenina ave.  47,
61103 Kharkov, Ukraine}

\author{Yu.\ V.\ Paderno, A.\ N.\ Bykov, V.\ N.\ Paderno, and V.\ I.\
Lyashenko}

\address{ I. N. Frantsevich Institute of Problems in Materials science,
National Academy of Sciences of Ukraine, Krzhizhanovskogo str. 3, 03142
Kiev, Ukraine}

\begin{abstract}

The sound velocity is measured in polycrystalline MgB$_{2}$ synthesized 
from the elements, and the bulk modulus and the 
Debye temperature are calculated. The conversion of an elastic wave
into electromagnetic radiation is investigated in the mixed state.
The dynamic parameters of the vortex lattice are estimated.

\end{abstract}
\maketitle
\section*{SYNTHESIS AND REAL STRUCTURE}

Magnesium diboride powder was obtained by synthesis from the elements Mg
(98\% pure) and B (99.5\% pure) in an argon medium. The powder was
sintered in a high-pressure chamber of the anvil-cell type
with a lenticular cell. The necessary compression was provided by a
hydraulic press with a 6300 kN force. The powder was heated
to the required temperature, not by the conventional indirect
heating,\cite{1} but by passing a current through the cylindrical
samples to be sintered, which gave a more uniform temperature
distribution over the volume of the sample. The preliminary compaction of
the powder was done by a cold two-sided pressing at pressures of up to
1.3 GPa. The pressed sample was placed in a capsule of graphitic
boron nitride, which was mounted in the working channel of the
high-pressure chamber. After completion of the sintering cycle the
material was cooled under pressure, and then the pressure was reduced
to atmospheric. The rate of buildup and relief of the pressure was 0.5
GPa/s. During the heating and cooling stages the temperature was
changed at a rate of 100 deg/s. The chosen construction of the cell of
the high-pressure chamber provided for hydrostatic compression of the
material to be sintered to pressures of up to 4.5 GPa and heating to
temperatures of up to 2000 K with holds of up to 300 s.

According to the data of an x-ray analysis the samples contained magnesium
diboride MgB$_{2}$ and an insignificant amount of manganese oxide MgO.

Microstructural and fractographic studies of the samples were carried
out on a Camebax CX-50 setup. The x-ray microspectral analysis
established the presence of the characteristic lines of magnesium,
boron, and oxygen.

A study of the real structure of the samples was done by the thin-foil
and diffraction microanalysis methods with the use of a P\'{E}M-U
transmission electron microscope.

The results of the study showed that the samples are characterized by
appreciable scatter in the grain sizes: in addition to very fine
particles ($d\approx 0.1$ $\mu$m) there are also rather large ones (up
to 30 $\mu$m), in agreement with the results of Ref.\ \onlinecite{2}.

\begin{figure}
\includegraphics[width=7.5cm,angle=0]{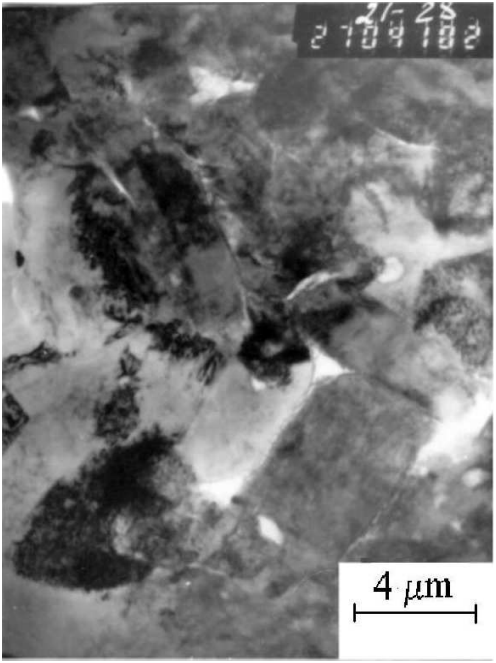} \caption[]{Typical
picture of the real structure of the MgB$_{2}$ sample,
demonstrating the substantial inhomogeneity of the structure and
the presence of appreciable scatter in the size of the grains. }
\label{f1}
\end{figure}

Some of the large grains are highly fragmented, apparently as a result
of the deformation during the preparation of the sample. It was noticed
that those grains had a high density of defects, contained fine
microcracks and pores, and had large dislocation pileups (Fig.\
\ref{f1}). The data of electron microanalysis also attest to strong
fragmentation: most of the reflections were smeared into arcs.

In addition to the fragmented grains, there were also rather large
($d>5$ $\mu$m) low-defect grains having a cubic configuration, which is
characteristic of the compound MgO, which crystallizes in a cubic
structure.

In certain parts of the structure, in particular, near pores and along
the boundaries of individual, clearly faceted grains, the presence of
interlayers of an amorphous phase, similar to that observed in Ref.\
\onlinecite{3}, was noted.

\section*{SOUND VELOCITY}

As far as we know, no experimental results from measurements of the
sound velocity in MgB$_{2}$ have been published yet. In imperfect
samples the most reliable data on the sound velocity can be obtained
only by analysis of the first signal, which has passed through the
sample along the shortest path. Interconversion of different modes at
inhomogeneities (including at grain boundaries) makes the more
distant reflections less suitable for measurements, as does to use of
some version of a resonance methods.

\begin{figure}
\includegraphics[width=7.5cm,angle=0]{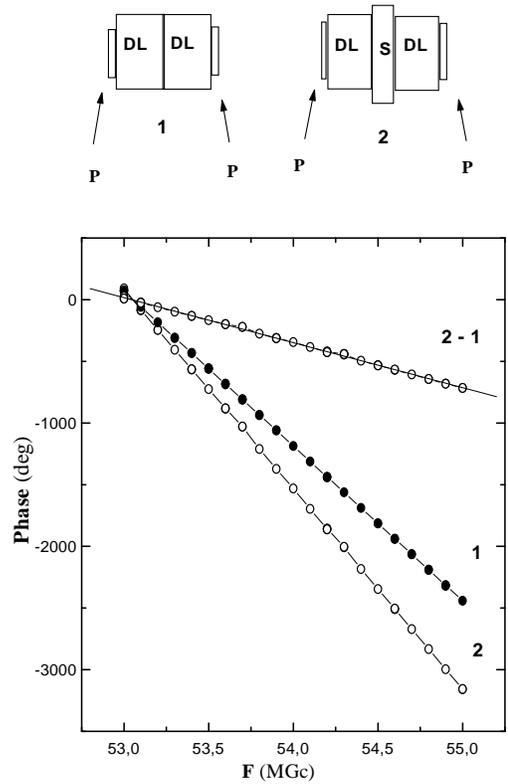}
\caption[]{ An example of the determination of the sound velocity
$V$ (transverse sound in MgB${_2}$) .The top - the symplified
scheme of the experiment . Here P - piezoelectric transduser, DL -
delay line, S - the sample. Lower : line 1 - the phase--frequency
characteristic (PFC) of two DL only ; line 2 - PFC of two DL and S; 
line 2 - 1 - the difference between PFC2 and PFC1. 
Sound velocity is determinated by equation $V = 360 l / A$ , 
$A$ - the slope of the line 2 - 1 $( 365 \cdot {
10^{-6} }Deg \cdot sec )$; $l$ - the sample length (0.49 cm). }
\label{f2}
\end{figure}

The method used in this paper was modified somewhat from that described
in Ref.\ \onlinecite{4}. In essence it consists in the measurement, in
a fixed frequency interval, of the phase--frequency characteristics of
two acoustic lines, consisting of delay lines with a sample
between them and delay lines without a sample (Fig.\ \ref{f2}).
The difference of the
phase--frequency curves in the absence of interference distortions in
the sample is a straight line with a slope determined by the sound
velocity that is sought (Fig.\ \ref{f2}). For millimeter sample lengths the error of
measurement of the absolute value of the frequency at frequencies of
${\sim }55$ MHz lies in the range 0.2--0.3\% at a signal-to-noise ratio
of ${\sim }3$.

The velocity values measured at $T = 77$ K (acoustic path length ${\sim
}5$ mm), viz., $V_{t} = 4.83\times 10^{5}$ cm/s and $V_{l} = 8.2\times
10^{5}$ cm/s are only effective values, since defects (pores,
microcracks) decrease the sound propagation velocity.\cite{5} The
MgB$_{2}$ sample under study had a well-defined geometric shape, making
it possible to estimate its porosity (${\sim }0.13)$ rather accurately
by a simple weighing. Using the approach proposed in Ref.\
\onlinecite{5}, we obtained the corrected values of the velocities:
$V_{t} = 5.12\times 10^{5}$ cm/s and $V_{l} = 8.76\times 10^{5}$ cm/s.
Finally, the limitations imposed in the porous-medium model make this
procedure much more uncertain than the error of measurement of the
effective values of the velocity. The values obtained are considerably
lower than the calculated values,\cite{6} although it may be that the
correction introduced is not complete, since it does not make allowance
for the presence of microcracks. Starting from the corrected
values of the velocities and the x-ray density, we calculated the
bulk modulus ($B_0 = 110$ GPa) and Debye temperature ($\Theta _{D} =
787$ K). The latter value is close to that corresponding to the
quantities obtained from heat-capacity measurements ($\Theta _{D} =
746$--$800$ K; see the literature cited in Ref.\ \onlinecite{6}).

\section*{DYNAMICS OF THE VORTEX PHASE}

The critical temperature ($T_{c} = 37$ K) and transition width ($\delta
T_{c}\approx 2$ K) are determined from the measurements of the
high-frequency magnetic susceptibility.

To estimate the dynamical parameters of the vortex lattice we have for
the first time used the method proposed in Ref.\ \onlinecite{7},
wherein they are extracted from data on the conversion of transverse
sound into an electromagnetic field. The sound wave vector ${\bf q}$
was oriented parallel to the external magnetic field ${\bf H}$ and
orthogonal to the MgB$_{2}$--free-space interface. The electromagnetic
radiation was registered by a coil oriented with the plane of its
windings also parallel to ${\bf H}$. The amplitude and phase of the
Hall component of the electromagnetic field were measured (the electric
field vector ${\bf E}_{H}$ was orthogonal to the displacement vector
{\bf u} in the sound wave). Such experiments had been done before
(only the amplitude ${\bf E}_{H}$ had been recorded) on either
low-temperature\cite{8} or high-temperature\cite{9} superconductors, and a simplified
theoretical approach to their description is proposed in Ref.\
\onlinecite{10}. However, the relations obtained in Ref.\
\onlinecite{10} do not admit a correct passage to the limit of the
normal state, as will be necessary in order to obtain quantitative
information (see below).

A more detailed analysis in the framework of the continuum
approximation and the two-fluid model\cite{7} leads to the relation
\begin{equation}
E_{H} = { \frac {1} {c} } B{ \dot u} \biggl(
{ \frac {k^2_L + k^2_{ns}}
{q^2 + k^2_L + k^2_{ns}} }\biggr)
\Biggl(
{ \frac { \alpha^{ \ast} }
{ \alpha^{ \ast} + q^2 { \frac {B^2} {4 \pi}}
{ \frac {k^2_L} {q^2 + k^2_L +k^2_{ns}} } }
}
\Biggr) \ ,
\end{equation}

Here $B$ is the induction in the sample, which is practically equal to
the external field $H$ except in a narrow field region around $H_{c1}$.
The wave numbers appearing in (1) have the following model dependences
on temperature and field:
\begin{center}
$k^2_L = \lambda^{-2} (1 - t^4)(1 - b)$ ,
$k^2_{ns} = 2i \delta^{-2}(1 -(1 -t^4)(1 - b))$ ,
\end{center}
where $\lambda $ is the low-temperature penetration depth in the
Meissner phase, $\delta $ is the skin penetration depth ($\delta ^{-2}
= 2\pi \omega \sigma _{n}/c^{2}$; $\sigma _{n}$ is the conductivity of
the normal metal), $t = T/T_{c}$, and $b = B/B_{c2}$ ($T_{c}$ and
$B_{c2}$ are the critical values of the temperature and magnetic
field).\\

The complex quantity $\alpha ^{\ast } = i\omega \eta +\alpha _{L}$
($\eta $ is the coefficient of friction and $\alpha _{L}$ is the
Labusch ``spring'' parameter) describes the dynamics of the vortex
lattice, which vibrates during the propagation of an elastic wave.
Bardeen and Stephen\cite{11} gave a simple estimate for $\eta $: $\eta
= BB_{c2}\sigma _{n}/c^{2}$. It may be noted that $\alpha
_{L}\rightarrow 0$ for $b\rightarrow  1$.

For $b = 1$ we have $k_{L} = 0$ and $k_{ns}^{2} = 2i\delta ^{-2}$, and
Eq.\ (1) reduces to the known expression for a normal metal in the
local limit.\cite{12} In the normal state the phase of $E_{H}$ leads
$\dot{u}$ by $\varphi _0 = {\rm arctan}(q^{2}\delta ^{2}/2)$, and
$|E_{H}c/\dot{u}|$ varies linearly with $H$, with a slope
$(1+\tan^{2}\varphi _0)^{-0.5}$. According to published data,\cite{13}
for MgB$_{2}$ the penetration depth $\lambda $ lies in the range
800--2000 \AA{}. At the frequencies used here $q\sim \delta ^{-1}\sim
10^{3}$, and therefore for $(1-b)>10^{-2}$ and at practically any
temperature we have $k_{L} \gg q,|k_{ns}|$, and relation (1) reduces to
\begin{equation}
E_{H} = { \frac {1} {c} } B{ \dot u}
\Biggl(
{ \frac { \alpha^{ \ast} }
{ \alpha^{ \ast} + q^2 { \frac {B^2} {4 \pi}}}}
\Biggr) \ ,
\end{equation}
which is suitable all the way down to the Meissner state. We
emphasize that the factor multiplying $q^{2}$ in Eq.\ (2), unlike the
situation in Eq.\ (1), is not identical to the bending modulus $C_{44}$
of the vortex lattice, since the latter should soften as $H_{c2}$ is
approached.\cite{14} For $(1-b) \ll 1$ ($\omega \eta \approx
2B_{c2}^{2}/4\pi \delta ^{2}$, $\alpha _{L}\approx 0$,
$k_{ns}^{2}\approx 2i\delta ^{-2})$ both Eqs.\ (2) and (1) reduce to
the corresponding relation for the normal metal at any value of
$k_{L}^{2}$. This means that even in an ideally homogeneous sample the
electric field $E_{H}$ changes continuously (without a jump) at the
transition of $H$ through $H_{c2}$. As $H$ is decreased further, the
evolution of $E_{H}$ depends on the relationship between $\omega \eta
$, $\alpha _{L}$, and the elastic modulus $C_{44} = B^{2}/4\pi $ of
the vortex lattice. Obviously, sooner or later the situation $|\alpha
^{\ast }| \gg C_{44}$ will arise, and in that case the phase of $E_{H}$
will coincide with the phase of $\dot{u}$, and $|E_{H}c/\dot{u}|$
will also vary linearly with the field, but now with a unit slope. Thus
in the case $\lambda ^{-2} \gg q^{2},\delta ^{-2}$ relation (2)
describes the behavior of $E_{H}$ in the entire field interval from
$H_{c2}$ to $H_{c1}$.

\begin{figure}
\includegraphics[width=7.5cm,angle=0]{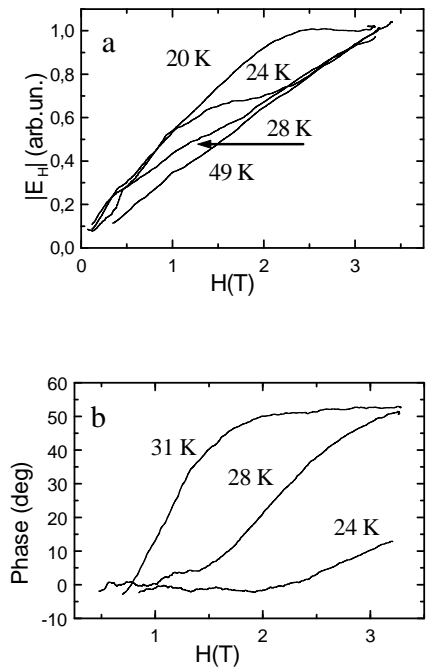}
\caption[]{Field dependence of the phase (a) and amplitude (b) of
the Hall component of the radiated electromagnetic field at
different temperatures. } \label{f3}
\end{figure}

The measured $E_{H}$ curves are presented in Fig.\ \ref{f3}. In view of
the limited range of magnetic fields accessible in our measurements,
informative experiments could be done only at rather high temperatures.
We see that the phase of the signal at the transition of $H$ through
$H_{c2}$ changes by approximately $\varphi _0\approx 50^{\circ}$ (Fig.\
\ref{f3}a). The ratio of the slopes (${\sim }1.6)$ in the regions of
linear variation of $|E_{H}|$ is in good agreement with the value that
follows from our analysis: $\sqrt{1+\tan^{2}\varphi _{0}}$ (Fig.\
\ref{f3}b). Knowing $\varphi _0$, we can estimate the depth of the skin
layer in the normal phase at the frequencies used here ($\delta \approx
2.3\times 10^{-3}$ cm) and the resistivity at $T \geq T_{c}$ ($\rho
_{n} = 12.5$ $\mu \Omega \cdot {\rm cm})$.

\begin{figure}
\includegraphics[width=7.5cm,angle=0]{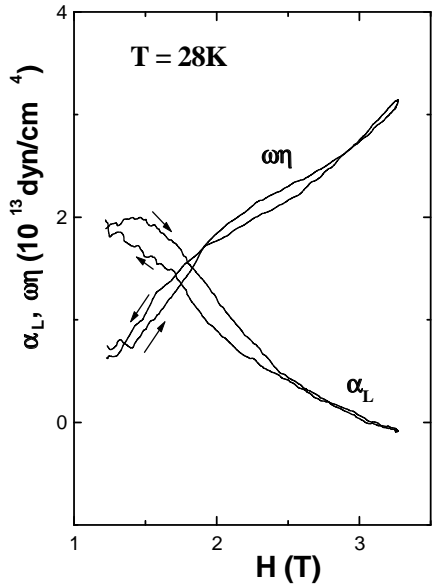}
\caption[]{Reconstruction of the coefficient of friction $\omega
\eta $ and Labusch parameter $\alpha _{L}$. } \label{f4}
\end{figure}

Relation (2) can easily be inverted to extract the real and imaginary
parts of $\alpha ^{\ast }$ from experimental data without any
additional assumptions. Of course, this can be done in the case when
one is able to track the variation of both the modulus and phase of
$E_{H}$ in the accessible magnetic field region. It is also obvious
that this procedure is doable only in that field region where the
amplitude and phase of $E_{H}$ deviate noticeably from their limiting
values. The result of such a construction is shown in Fig.\ \ref{f4}.

As expected, $\eta $ varies practically linearly with applied field.
For a known $\sigma _{n}$, the slope of this relation is determined, in
accordance with Ref.\ \onlinecite{11}, by the value of $H_{c2}$.
Extrapolation of the data in Fig.\ \ref{f4} by a straight line passing
through the origin gave $H_{c2}\approx 3.2$ T, practically equal to the
field that determines the beginning of the change in the phase of
$E_{H}$ at $T = 28$ K (Fig.\ \ref{f3}a). This means that in the
MgB$_{2}$ sample under study, $H_{c2}$ is limited by the condition of
overlap of the cores of the vortices and not by the Clogston
paramagnetic limit.\cite{12}

It is seen from the results presented in Fig.\ \ref{f3} that the
main change in the phase of $E_{H}$ is practically finished by the
start of the deviation of $|E_{H}|$ from the linear dependence
characterizing the normal phase. This means that in this stage
the inequality $\omega \eta <\alpha _{L} \leq q^{2}C_{44}$ holds, and
$\alpha _{L}$ can be estimated either by inverting relation (2) with
$\alpha ^{\ast }\approx \alpha _{L}$ or simply from the relation
$\alpha _{L}\approx q^{2}H_{m}^{2}/4\pi $, where $H_{m}$ is the
characteristic field at which the transition of $|E_{H}|$ from one
linear trend to another occurs. In particular, for $T = 20$ K an
estimate of the Labusch parameter gives $\alpha _{L}\sim 4\times
10^{13}$ dyn/cm$^{4}$. Using the order-of-magnitude relation $\alpha
_{L}\approx BI_{c}/cr$, where $I_{c}$ is the critical current and $r$
is the characteristic spatial scale of the pinning potential, which for
$H \gg H_{c1}$ is close to the period of the vortex structure, we
obtain $I_{c}\approx 3\times 10^{4}$ A/cm$^{2}$ ($T = 20$ K, $H\approx
3$ T).

This study was supported by the Government Foundation for Basic
Research of the Ministry of Education and Science of Ukraine.

This paper was pubilshed in Low Temp. Phys. No. 3, March 2002, p.270-274;

\noindent

\end{document}